\input{aipcheck}

\documentclass[final
    ,final            % use final for the camera ready runs
  ]
  {aipproc}

\layoutstyle{6x9}

\begin{document}

\title{Mirror dark matter interpretation of the DAMA/Libra 
annual modulation signal
%\footnote{Talk given at DSU09}  
}

\classification{95.36.+d,   12.60.-i}
\keywords      {dark matter}

\author{R. Foot}{
  address={ School of Physics,University of Melbourne,Victoria 3010 Australia\\
Email: rfoot@unimelb.edu.au } }

\begin{abstract}
The DAMA/Libra experiment has recently confirmed the annual
modulation signal obtained in the earlier DAMA/NaI experiment,
providing strong evidence that they have actually detected dark matter.
We examine the implications of this experiment for the
mirror dark matter candidate. We show that mirror
dark matter successfully explains the latest DAMA/Libra
data, whilst also being consistent with the null
results of other direct detection experiments.

\end{abstract}

\maketitle

%%%%%%%%%%%%%%%%%%%%%%%%%%%%%%%%%%%%%%%%%%%%
%% MAINMATTER
%%%%%%%%%%%%%%%%%%%%%%%%%%%%%%%%%%%%%%%%%%%%

%\section{Introduction and the DAMA experiments}

I'm interested in talking about some recent developments in efforts aimed at 
the direct detection
of dark matter. There are a number of on-going experiments, but there
is one experiment which is particularly interesting because
it seems that they've actually detected dark matter.
This is the 
DAMA/Libra experiment\cite{damalibra} and its predecessor the DAMA/NaI experiment\cite{dama}.

These DAMA experiments eliminate the background by using the annual modulation 
signature. The idea\cite{idea} is very simple.
The interaction rate must vary periodically since it depends on the Earth's velocity, $v_E$, which 
modulates due to the Earth's motion around the Sun. That is,
\begin{eqnarray}
R(v_E) = R (v_{\odot}) + \left( \frac{\partial R}{ \partial v_E}\right)_{v_{\odot}} \Delta v_E \cos \omega (t-t_0)
\end{eqnarray}
where $\Delta v_E \simeq 15$ km/s, $\omega \equiv  2\pi/T$ ($T = 1$ year) and $t_0 = 152.5$ days (from
astronomical data).
The phase and period are both predicted! This gives a strong systematic check on their results.
Such an annual modulation was found\cite{damalibra} at the $8.2\sigma$ Confidence level, with $T, t_0$ measured to be:
\begin{eqnarray}
T &=& 0.998 \pm 0.003 \ {\rm year} \nonumber \\
t_0 &=& 144 \pm 8 \ {\rm day}
\end{eqnarray}
Clearly, both the period and phase are consistent with the theoretical expectations of halo dark matter.
These are strong reasons to believe that the DAMA people have detected dark matter.
The data, together with the cosine prediction is given in figure 1.

%%%%%%%%%%%%%%%%%%%%%%%%%%%%%%%%%%%%%%%%%%%%
%% Sample figure:
%%
%% The option [height=...] scales the picture to the given height,
%% without it it would be printed at its nominal size
%%%%%%%%%%%%%%%%%%%%%%%%%%%%%%%%%%%%%%%%%%%%

\begin{figure}
  \includegraphics[height=.5\textheight,angle=270]{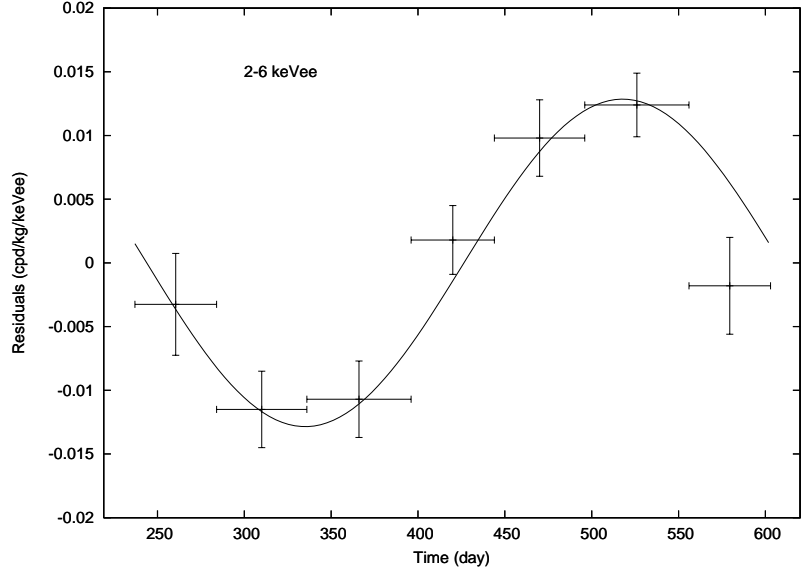}
\caption{DAMA/Libra annual modulation signal (in the 2-6 keV recoil energy region) 
together with the dark matter prediction.
Note that the initial time in this figure is August 7th.}
\end{figure}

%\section{Theory}

What type of theory could explain these results? As a first try, one could take 
a simple minded approach to the question of dark matter. One could imagine that the laws 
governing dark matter could
be identical to the laws governing ordinary matter. By this I mean that the dark matter
could belong to a sector which is an exact duplicate of the ordinary matter sector, so that
the Lagrangian is:
\begin{eqnarray}
{\cal L} = {\cal L}_{SM} (e, u, d, \gamma,...) + {\cal L}_{SM} (e', u', d', \gamma', ...)
\end{eqnarray}
Such a theory can also be motivated from a symmetry reason if left and right handed chiral
fields are interchanged in the extra sector. In this way space-time parity symmetry
can be realized as a symmetry of nature, and for this reason we call the particles in
the extra sector mirror particles. The standard model extended with 
a mirror sector was first studied in ref.\cite{flv} and shown to be a consistent extension
of the standard model 
(for a review and more complete list of references see ref.\cite{review}).

Observe that $P', e', He' $ etc would have masses identical to their ordinary matter
counterparts. They would also be stable and thus potentially dark matter candidates.
In order to make this type of theory consistent with the 
successful big bang nucleosynthesis (BBN) measurements and successful 
large scale structure (LSS), we need to assume
that in the early Universe, i.e. when the temperature was around 1 MeV and less,
that $T' < T$. In fact, some studies suggest that we require $T'/T \stackrel{<}{\sim} 0.3$
\cite{some}.

The idea as it stands doesn't explain any dark matter direct detection experiment since the ordinary
and mirror particles interact with each other only via gravity.
A relevant question is: can ordinary and mirror particles interact with each other non-gravitationally?
That is, can we add any interaction terms consistent with renormalizability and the symmetries of
the theory?
The answer is YES - but only two terms are possible\cite{flv}:
\begin{eqnarray}
{\cal L}_{mix} = \frac{\epsilon}{2} F^{\mu \nu} F'_{\mu \nu} + \lambda \phi^{\dagger}\phi \phi'^{\dagger} \phi' \ ,
\end{eqnarray}
where $F_{\mu \nu}$ ($F'_{\mu \nu}$) is the ordinary (mirror) $U(1)$ gauge boson field strength tensor and 
$\phi$ ($\phi'$) is the electroweak Higgs (mirror Higgs) field.
Both of these terms can lead to interactions between the ordinary and mirror particles. It turns out that
the coupling of the Higgs to protons and neutrons is too weak for the Higgs mirror Higgs quartic
term to be important in dark matter experiments. This leaves us with the photon mirror photon kinetic
mixing term, which it turns out, can do the job \cite{foot08,foot03}. 
In fact this term will lead to Rutherford-type elastic
scattering (figure 2) of ordinary nuclei off mirror nuclei (for the
relevant cross-section, form factors, and other technical details,
see ref.\cite{foot08,foot03}).
\begin{figure}
  \includegraphics[height=.2\textheight]{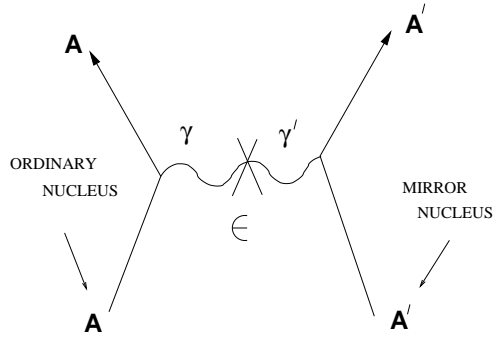}
\caption{Rutherford elastic scattering of mirror nuclei off ordinary nuclei induced
by photon - mirror photon
kinetic mixing of strength $\epsilon$.}
\end{figure}

To make contact with the direct detection experiments we need to know something about the chemical
composition and distribution of particles in the galactic halo.
\vskip 0.3cm
\noindent
{\it Chemical composition:}
\vskip 0.2cm
\noindent
$H', He'$ are expected to be produced in the early Universe, 
and in fact we expect a larger $He'$ mass fraction, $Y_{He'} > Y_{He}$, if $T' < T$.
Elements heavier than $He'$, such as $O', Ne'....$ are expected to be produced in mirror
stars.
\vskip 0.3cm
\noindent
{\it Distribution:}
\vskip 0.2cm
\noindent
To explain the rotation curves in spiral galaxies, we know that the dark matter needs to be
roughly spherically distributed in spiral galaxies. Given the upper limit on compact star
sized objects in the halo from MACHO searchers (roughly $f_{macho} \stackrel{<}{\sim} 0.2-0.3$
depending on the assumptions), we then expect the mirror particles to be 
distributed predominately
as a gaseous spherical halo surrounding the collapsed disk of ordinary matter\cite{fv}.
A dissipative dark matter candidate like mirror matter can only survive in an extended spherical
distribution without collapsing
if there is a substantial heating mechanism to replace the energy lost due to radiative cooling.
It turns out, that there is such a heating mechanism, which I will comment on, at the end of this talk.

Anyway, the gas of mirror particles will have a Maxwellian velocity distribution,
\begin{eqnarray}
f[i] &=& e^{-\frac{1}{2} m_i v^2/T}  \nonumber \\  
 &=& e^{-v^2/v_0^2[i]} 
\end{eqnarray}
where the index $i$  labels the particle type [$i=e', P', He', O', ...$]. 
In the non-interacting one species WIMP case, $T = \frac{1}{2} m v_{rot}^2$, (where $v_{rot} \sim 250$ km/s is
the rotational velocity in the Milky Way). In the mirror matter case
where we have several particle species with significant self interactions, we must examine the condition for
hydrostatic equilibrium, which balances the central gravitational attraction with the pressure
gradient:
\begin{eqnarray}
\frac{dP}{dr} = - \rho g
\end{eqnarray}
where
\begin{eqnarray}
P = \sum n_i T, \ \rho = \sum m_i n_i, \  g = \frac{G}{r^2}\int^r \rho dV = {v_{rot}^2 \over r}
\ .
\end{eqnarray}
Here, $n_i, \rho, T$
are the number density of species $i$ ($i=e', P', He',...$), mass density and Temperature,
which is assumed to be isothermal for simplicity.
It is straightforward to solve this equation, leading to\cite{foot03}:
\begin{eqnarray}
T = \frac{1}{2} \overline{m} v_{rot}^2
\end{eqnarray}
where $\overline{m} = \sum m_i n_i/\sum n_i$ is the average mass of the particles in the halo. 
Thus we see that the velocity dispersion of the particles in the dark matter halo depends
on the particular particle species and satisfies:
\begin{eqnarray}
v_0^2 [i] = v_{rot}^2 \frac{\overline{m}}{m_i}
\end{eqnarray}
Note that if $m_i \gg \overline{m}$, then $v_0^2[i] \ll v_{rot}^2$. That is, heavier mirror particles
will have a very narrow velocity distribution, which turns out to be a key feature in
explaining why DAMA sees a signal and the other experiments do not (within this framework).
In fact, given that the kinetic energy of halo mirror particles in the Earth's reference frame is 
$E = \frac{1}{2} mv^2_{rot} \sim keV$ for $m_i = m_{He}$, it follows that with DAMA's 2 keV recoil 
energy threshold that
they are indeed only sensitive to elements heavier than $He$, that is the sub dominant component, 
naively expected to be mirror oxygen $O'$,
and from the above, will have narrow velocity distribution $v_0^2 [O'] \ll v_{rot}^2$.

%\section{Comparison of theory and experiment}

Fixing the normalization of the annual modulation gives a measurement of $\epsilon$, 
which turns out to be $\epsilon \sim 10^{-9}$\cite{foot08}.
A value for $\epsilon$ of around this magnitude is consistent with experimental 
and astrophysical constraints (for
a review, see. ref.\cite{footrev}), and also consistent with early Universe 
cosmology bounds (successful BBN and LSS)\cite{paolo}.

Importantly, the DAMA/Libra experiment provides new information because they 
have enough data to begin to examine
the dependence of their signal on the recoil energy of the target nuclei, 
which is something that they measure.
We define the annual modulation amplitude $S^m$ by:
\begin{eqnarray}
{dR(v_E) \over dE_R} &=& {dR (v_{\odot})\over dE_R}  + 
\left( \frac{\partial dR/dE_R}{ \partial v_E}\right)_{v_{\odot}} \Delta v_E \cos \omega (t-t_0)
\nonumber \\
 &=& {dR (v_{\odot})\over dE_R}  + 
S^m \cos \omega (t-t_0) \ .
\end{eqnarray}
How do we expect $S^m$ to vary with $E_R$? For large $E_R \gg \frac{1}{2} m v_{rot}^2$, 
only particles in the tail of the Maxwellian distribution could produce 
such a large $E_R$ scattering event in the Earth based detector.
On June 2nd, when $v_E$, 
is a maximum, we expect more large $E_R$ scattering events, since there will be a 
lot more mirror particles which
are energetic enough (in Earth's reference frame) to produce a given large $E_R$ scattering.
Thus, in the large $E_R$ region, we expect $S^m$ to be positive and tending to zero as
$E_R \to \infty$.   
At low $E_R \ll \frac{1}{2} mv_{rot}^2$, all the particles
in the halo are energetic enough to produce a scattering event, but since $d\sigma/dE_R \propto 1/v^2$, 
we expect less events when $v_E$ is a maximum, so that $S^m < 0$ at low $E_R$.
To summarize, we expect $S^m$ to have an $E_R$ dependence which goes to 0 as $E_R \to \infty$,
rises to a peak as $E_R$ moves from infinity towards the body of the distribution, and changes sign 
at low $E_R$.
\begin{figure}
  \includegraphics[height=.5\textheight,angle=270]{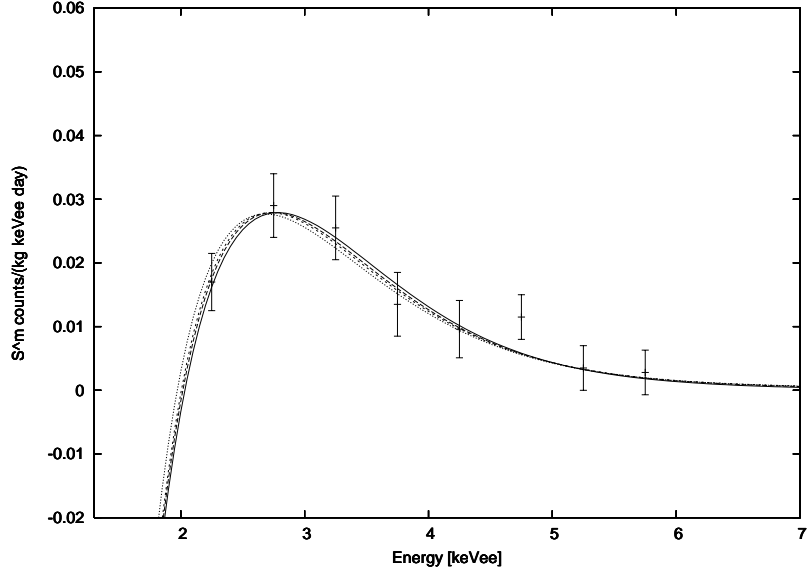}
\caption{
Energy dependence of the cosine modulation amplitude, $S^m$, for four illustrative cases:
$A'=Si', v_{rot}=170$ km/s (solid line)
$A'=Mg', v_{rot}=195$ km/s (long-dashed line),
$A'=Ne', v_{rot}=230$ km/s (short-dashed line),
$A'=O', v_{rot}=280$ km/s (dotted line).
In each case $\epsilon$ is fixed so that the mean amplitude is 0.0129 cpd/kg/keVee.
Also shown are the DAMA/NaI \& DAMA/LIBRA combined data from figure 9 of ref.\cite{damalibra}.
This figure assumes that the mass of the halo is dominated by $He'$.}

\end{figure}

Note that the position of the peak, which is essentially given by kinematics, 
will give a measure of the
mass of the dark matter particle. 
In ref.\cite{foot08} I have done the analysis, taking into account all the required things, 
such as energy resolution
of the detector, quenching factors etc. 
Assuming a halo dominated in mass by $H'/He'$ with an $A'$ subcomponent (where $A' \sim O'$ is
the expectation), I find a predicted shape (see figure 3) for $S^m$
which agrees nicely with the shape measured by DAMA/Libra.    
This is a significant test of the theory.

\begin{figure}
  \includegraphics[height=.5\textheight,angle=270]{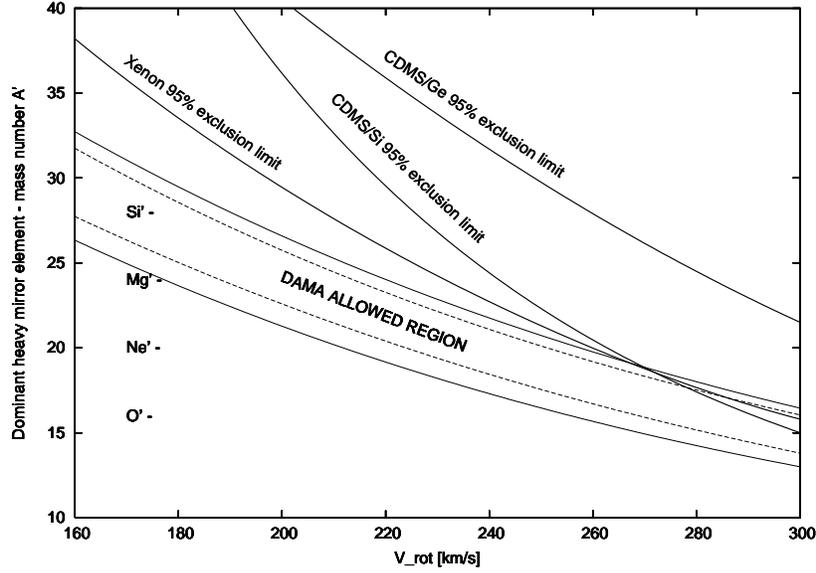}
\caption{
DAMA/Libra allowed region together with the 95\% C.L. exclusion limits
from the XENON10\cite{xenon}, CDMS/Ge and CDMS/Si \cite{cdms} experiments 
(the regions above the exclusion
contours  are the disfavoured region).
}
\end{figure}

Allowing for a range of possible $v_{rot}$ (most recent 
astronomical data suggest\cite{reid} $v_{rot} = 254 $ km/s),
leaves a band of allowed parameter space, shown in figure 4. 
Assuming $v_{rot} = 254 \pm 30$ km/s, 
a galactic halo consisting of a dominant $H',He'$ with $A'$ subcomponent can fit
the data for $A'$ with mass $18\pm 4$ GeV, which is compatible with the naive
expectation of $O'$ (given that oxygen is the most abundant ordinary element
after H/He).
It turns out that ALL of this parameter space is
consistent with the null results of the other experiments. The 95\% C.L. limits of the most 
sensitive of these null experiments, are also indicated.

These results are in sharp contrast to  
popular models involving weakly interacting massive particles (WIMPs).
Elastic scattering of standard
WIMPs give a very poor fit to the data\cite{wimp1}.
The basic reasons as to why elastic scattering with mirror dark matter works, and 
elastic scattering of standard WIMPs doesn't has to 
do with their basic differences:
a) the velocity distribution of $O'$ dark matter is much narrower than for WIMPs,
$v_0^2 [O'] \ll v_{rot}^2$, while for standard WIMPs $v_0^2 = v_{rot}^2$, and
b) Rutherford scattering has a cross section $d\sigma/dE_R \propto 1/E_R^2$,
while for standard WIMPs $d\sigma/dE_R$ is $E_R$ independent (excepting
the $E_R$ dependence on the form factors).
I would also argue that mirror dark matter is simpler and more elegant than
models with
standard WIMPs, so in a sense its not surprising that experiments have come out
in favour of mirror dark matter, but of course I might be biased!

%\section{Concluding remarks}
%\vskip 0.3cm
%\noindent

This mirror matter interpretation of the dark matter detection 
experiments  will be tested further by DAMA -  as they collect more data 
their statistical error will reduce.
Furthermore the DAMA people plan to upgrade
their experiment replacing their PMTs with the aim of lowering their $E_R$ threshold. This will be
particularly useful, as they should see the change in sign of their modulation $S^m$, predicted
in figure 3.
% \vskip 0.3cm
%\noindent
Experiments like XENON10\cite{xenon} and CDMS\cite{cdms} have $E_R$ threshold too high 
to see the same 
mirror dark matter component which DAMA
is detecting. They could still find a positive signal if there is a heavier component, such as an $Fe'$ 
component. Such a component should be there at some level.
%\vskip 0.3cm
%\noindent

The inferred value of $\epsilon \sim 10^{-9}$ seems quite interesting for supernova
physics.
In particular for such values of $\epsilon$ about half of the total energy emitted in Supernova
explosions will be in the form of light mirror particles ($\nu'_{e,\mu,\tau}$, $e'^{\pm}, \gamma'$)
\cite{zurab}.
This implies a heating of the halo (principally due to the $e'^{\pm}$ component), of around:
\begin{eqnarray}
L^{SN}_{heat-in} \sim \frac{1}{2} \times 3\times 10^{53}\ erg 
{1 \over 100\ years} \sim 10^{44}\ {\rm erg/s \ for \ Milky\ Way
}
\end{eqnarray}
It turns out that this matches (to within uncertainties) the energy lost from the halo due to 
radiative cooling\cite{fv}:
\begin{eqnarray}
L^{halo}_{energy-out} = \Lambda \int_{R_1} n^2_{e'} 4\pi r^2 dr \sim 10^{44} \ {\rm erg/s \ for
\ Milky \ Way}.
\end{eqnarray}
In other words, a gaseous mirror particle halo can survive without collapsing because
the energy lost due to dissipative interactions is replaced by the energy from ordinary supernova
explosions. Presumably there is some detailed dynamical reasons maintaining this balance,
which of course, may be difficult to elucidate due to the complexity of that particular problem.

%%%%%%%%%%%%%%%%%%%%%%%%%%%%%%%%%%%%%%%%%%%%%%%%
%% BACKMATTER
%%%%%%%%%%%%%%%%%%%%%%%%%%%%%%%%%%%%%%%%%%%%%%%%

\begin{theacknowledgments}
This work was supported by the Australian Research Council.

\end{theacknowledgments}

\end{document}